\documentclass[aps,pra,twocolumn,groupedaddress,floatfix,twoside]{revtex4}

\usepackage{graphicx}
\usepackage{dcolumn} 
\usepackage{bm}

\begin{document}

\title{A molecular Bose-Einstein condensate emerges from a Fermi sea}

\author{Markus Greiner}
\email[Email: ]{markus.greiner@colorado.edu}
\author{Cindy A. Regal}
\author{Deborah S. Jin}
\altaffiliation[]{Quantum Physics Division, National Institute of
Standards and Technology.}

\affiliation{JILA, National Institute of Standards and Technology
and Department of Physics, University of Colorado, Boulder, CO
80309-0440, USA}

\date{November 3, 2003}

\begin{abstract}
{The realization of fermionic superfluidity in a dilute gas of
atoms, analogous to superconductivity in metals, is a
long-standing goal of ultracold gas research. Beyond being a new
example of this fascinating quantum phenomenon, fermionic
superfluidity in an atomic gas holds the promise of adjustable
interactions and the ability to tune continuously from BCS-type
superfluidity to Bose-Einstein condensation (BEC). This crossover
between BCS superfluidity of correlated atom pairs in momentum
space and BEC of local pairs has long been of theoretical
interest, motivated in part by the discovery of high $T_{c}$
superconductors.$^{1 - 9}$ In atomic Fermi gas experiments
superfluidity has not yet been demonstrated; however recent
experiments have made remarkable progress toward this goal.
Starting from an ultracold Fermi gas experimenters have used
Feshbach resonances to reversibly create molecules, i.e. composite
bosons consisting of local fermion pairs.$^{10 - 13}$ Furthermore,
the experiments have shown that the resulting diatomic molecules
can have surprisingly long lifetimes.$^{11 - 14}$ Here we report
the conversion of a Fermi sea of atoms into a molecular BEC. In
addition to being the first molecular condensate in thermal
equilibrium, this BEC represents one extreme of the predicted
BCS-BEC continuum.}
\end{abstract}

\maketitle

 The basic idea behind this experiment is to start with
a Fermi gas that has been evaporatively cooled to a high degree of
quantum degeneracy and adiabatically create molecules with a
magnetic-field sweep across the Feshbach resonance. If the
molecule creation conserves entropy and the initial atom gas is at
sufficiently low temperature $T$ compared to the Fermi temperature
$T_{F}$, then the result should be a molecular sample with a
significant condensate fraction.$^{12,15}$ With a relatively slow
sweep of an applied magnetic field that converts most of the
fermionic atoms into bosonic molecules and an initial atomic gas
below $T/T_{F}=0.17$, we observe a molecular condensate in
time-of-flight absorption images taken immediately following the
magnetic-field sweep. Strikingly, the molecular condensate is not
formed by any active cooling of the molecules, but rather merely
by traversing the predicted BCS-BEC crossover regime.
\begin{figure}
\includegraphics[width=\linewidth]{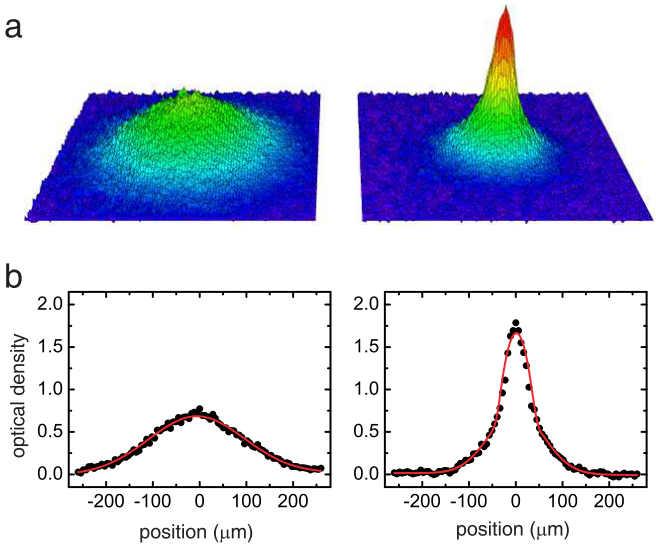}
\caption{Time-of-flight images of the molecular cloud along the
axial direction after 20~ms of free expansion for a temperature
above and below the critical temperature for Bose-Einstein
condensation. \textbf{a}, Surface plot of the optical density for
a molecule sample created by applying a magnetic-field sweep to an
atomic Fermi gas with an initial temperature of 0.19 $T_{F}$
(0.06~$T_{F})$ for the left (right) picture. Here the radial
trapping frequency of the optical trap was 350 Hz (260 Hz). When
we start with the lower initial temperature of the fermionic atoms
(right) and adiabatically ramp across the Feshbach resonance from
$B$~=~202.78 G to 201.54 G in 10~ms, the molecules form a
Bose-Einstein condensate. The interparticle interaction during
expansion was reduced by rapidly moving the magnetic field 4~G
further away from the Feshbach resonance. In this measurement the
total molecule number was 470,000 (200,000) for the left (right)
picture. The surface plots are the averages of 10 images.
\textbf{b}, Cross sections through images corresponding to the
parameters given above (dots), along with bimodal surface fits
(lines). The fits yield no condensate fraction and a temperature
of $T$ = 0.90~$T_{c}$ for the left graph, and a 12{\%} condensate
fraction and a temperature of the thermal component of $T$ = 0.49
$T_{c}$ for the right graph. Here, $T_{c}$ is the calculated
critical temperature for a noninteracting Bose-Einstein condensate
in thermal equilibrium.}
\end{figure}

Our experimental set-up and procedure used to cool a gas of
fermionic $^{40}$K atoms to quantum degenerate temperatures is
detailed in previous work.$^{16 - 18}$ In brief, after laser
cooling and trapping we evaporatively cool the atoms in a magnetic
trap. In order to realize s-wave collisions in the ultracold Fermi
gas we use a mixture of atoms in two different spin states. For
the final stage of evaporative cooling the atoms are loaded into
an optical dipole trap formed by a single far red-detuned laser
beam. The laser wavelength is $\lambda $~=~1064 nm and the beam is
focused to a waist of 15.5 $\mu $m. By lowering the depth of the
optical trap we evaporate the atomic gas to temperatures far below
the Fermi temperature $T_F = (6N\nu _r^2 \nu _z )^{1 / 3}h / k_B
$. Here $N$ is the particle number in each spin state,
\textit{$\nu $}$_{r}$ and \textit{$\nu $}$_{z}$ are the radial and
axial trap frequencies, $h$ is Planck's constant, and $k_{B }$is
Boltzmann's constant. For final radial trap frequencies between
$\nu _{r}$~=~430~Hz and 250~Hz and a fixed trap aspect ratio $\nu
_{r}$/$\nu _{z}$~=~79~$\pm $~15, we achieve temperatures between
0.36~$T_{F}$ and 0.04 $T_{F}$. All temperatures of the Fermi gas
given in this work are determined through surface fits to
time-of-flight absorption images.$^{17}$

For this work we use a Feshbach resonance, which occurs when the
energy of a quasibound molecular state becomes equal to the energy
of two free atoms.$^{19}$ The magnetic-field dependence of the
resonance allows for precise tuning of the atom-atom interaction
strength in an ultracold gas.$^{20,21}$ Moreover, time-dependent
magnetic fields can be used to reversibly convert atom pairs into
extremely weakly bound molecules.$^{10 - 13,22 - 28}$ The
particular resonance used here is located at a magnetic field
$B_{0}$~=~202.1~$\pm $~0.1~G and has a width of $w$ = 7.8~$\pm
$~0.6~G.$^{14,18}$ The resonance affects collisions between atoms
in the two lowest energy spin states $\left| {f = 9 / 2,m_f = - 7
/ 2} \right\rangle $ and $\left| {f = 9 / 2,m_f = - 9 / 2}
\right\rangle $, where $f$ denotes the total atomic angular
momentum and $m_{f}$ the magnetic quantum number.

To create bosonic molecules from the fermionic atoms, we first
prepare an equal mixture of atoms in the $m_{f}$~= -9/2 and
$m_{f}$~=~{\-}7/2 spin states at temperatures below quantum
degeneracy. Then we apply a time-dependent sweep of the magnetic
field starting above the Feshbach resonance value, where the atom
interactions are effectively attractive, and ending below the
resonance, where the atom interactions are effectively repulsive.
In contrast to our previous work$^{10}$ the magnetic-field sweep
is not only adiabatic with respect to the molecule creation rate,
but also slow with respect to the collision rate and the radial
trap frequency.$^{12}$ The magnetic field is typically ramped in
7~ms from $B$ =~202.78~G to either $B$~=~201.54 G or $B$~=~201.67
G. With this magnetic-field sweep across the Feshbach resonance we
convert between 78{\%} and 88{\%} of the atoms into molecules. To
a very good approximation these molecules have twice the
polarizability of the atoms$^{29}$ and therefore are confined in
the optical dipole trap with the same trapping frequency and twice
the trap depth of the atoms. The molecules, which are all in the
same internal quantum state, are highly vibrationally excited,
very large in spatial extent, and extremely weakly bound. For a
magnetic field 0.43 G below the Feshbach resonance ($B$=201.67 G)
the binding energy determined through a full coupled channels
calculation is 8 kHz $^{30}$ and the molecule size is $\sim
$1650~$a_{0}$, where $a_{0}$ is the Bohr radius. Here we estimate
the molecule size as $a/2$, where $a$ is the atom-atom scattering
length given by $a = 174a_0 [1 + w / (B_0 - B)]$. At this magnetic
field the molecule size is one order of magnitude smaller than the
calculated intermolecular distance.

A critical element of this experiment is that the lifetime of
these weakly bound molecules can be much longer than the typical
collision time in the gas and longer than the radial trapping
period.$^{11 - 14}$ In previous work we found that the
$^{40}$K$_{2}$ molecule lifetime increases dramatically near the
Feshbach resonance and reaches $\sim $100~ms at a magnetic field
0.43~G below the Feshbach resonance for a peak density of
$n_{pk}$~=~1.5~$\times $~10$^{13}$~cm$^{-3}$.$^{14}$ It is
predicted that this increased molecule lifetime only occurs for
dimers of fermionic atom pairs.$^{31}$ The relatively long
molecule lifetime near the Feshbach resonance allows the
atom/molecule mixture to achieve thermal equilibrium during the
magnetic-field sweep. Note however that the large aspect ratio of
the optical trap gives rise to a strongly anisotropic system. Thus
for the relatively short timescale of the experiments reported
here we may attain only local equilibrium in the axial
direction.$^{32}$

To study the resulting atom-molecule gas mixture after the
magnetic-field sweep, we measure the momentum distribution of both
the molecules and the residual atoms using time-of-flight
absorption imaging. After typically 10 to 20~ms of expansion we
apply a radio frequency (rf) pulse that dissociates the molecules
into free atoms in the $m_{f}$~=~{\-}5/2 and $m_{f}$~=~{\-}9/2
spin states.$^{10}$ Immediately after this rf dissociation pulse
we take a spin-selective absorption image. The rf pulse has a
duration of 140~$\mu $s and is detuned 50~kHz beyond the molecule
dissociation threshold where it does not affect the residual
unpaired atoms in the $m_{f}$~=~{\-}7/2 state. We selectively
detect the expanded molecule cloud by imaging atoms transferred by
the rf dissociation pulse into the previously unoccupied
$m_{f}$~=~{\-}5/2 state. Alternatively we can image only the
expanded atom cloud by detecting atoms in the $m_{f}$~=~{\-}7/2
spin state.

Close to the Feshbach resonance, the atoms and molecules are
strongly interacting with effectively repulsive interactions. The
scattering length for atom-molecule and molecule-molecule
collisions close to the Feshbach resonance has recently been
calculated by Petrov \textit{et al}. to be 1.2 $a$ and 0.6 $a$
respectively, where $a$ is the atom-atom scattering length.$^{31}$
During the initial stage of expansion the positive interaction
energy is converted into additional kinetic energy of the
expanding cloud. Therefore the measured momentum distribution is
very different from the original momentum distribution of the
trapped cloud. In order to reduce the effect of these interactions
on the molecule time-of-flight images we use the magnetic-field
Feshbach resonance to control the interparticle interaction
strength during expansion. We can significantly reduce the
interaction energy momentum kick by rapidly changing the magnetic
field before we switch off the optical trap for expansion. The
field is lowered typically by 4~G in 10~$\mu $s. At this magnetic
field further away from the resonance the atom{\-}atom scattering
length $a$ is reduced to $\sim $500~$a_{0}$. We find that this
magnetic-field jump results in a loss of typically 50{\%} of the
molecules, which we attribute to the reduced molecule lifetime
away from the Feshbach resonance.
\begin{figure}
\includegraphics[width=0.8\linewidth]{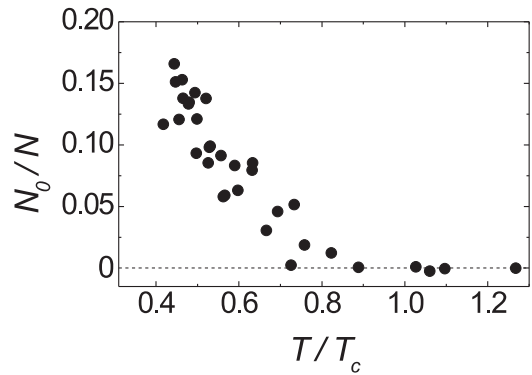}
\caption{Molecular condensate fraction $N_{0}/N$ versus the scaled
temperature $T/T_{c}$. The temperature of the molecules is varied
by changing the initial temperature of the fermionic atoms prior
to the formation of the molecules. All other parameters are
similar to the ones described in the caption of Fig. 1. We observe
the onset for Bose-Einstein condensation at a temperature of $
\sim $0.8 $T_{c}$.}
\end{figure}

Below an initial temperature of 0.17~$T_{F}$ we observe the sudden
onset of a pronounced bimodal momentum distribution for the
molecules. Figure 1 shows such a bimodal distribution for an
experiment starting with an initial temperature of 0.06 $T_{F }$;
for comparison we also show the resulting molecule momentum
distribution for an experiment starting at 0.19 $T_{F}$. The
bimodal momentum distribution is a striking indication that the
cloud of weakly bound molecules has undergone a phase transition
to a Bose-Einstein condensate.$^{33 - 35}$

In order to obtain thermodynamic information about the molecule
cloud we fit the momentum distribution with a two-component fit.
The fit function is the sum of an inverted parabola describing the
Thomas-Fermi momentum distribution of a bosonic condensate and a
Gaussian momentum distribution describing the non-condensed
component of the molecule cloud. In Fig. 2 the measured condensate
fraction is plotted as a function of the fitted temperature of the
thermal component in units of the critical temperature for an
ideal Bose gas $T_c = 0.94(N\nu _r^2 \nu _z )^{1 / 3}h / k_B $.
Here $N$ is the total number of molecules measured without
changing the magnetic-field for the expansion. Note that this
measurement may underestimate the original condensate fraction due
to loss of molecules during expansion. From Fig. 2 we determine an
actual critical temperature for the strongly interacting molecules
and for our trap geometry of 0.8~$\pm $~0.1~$T_{c}$. Such a
decrease of the critical temperature relative to the ideal gas
prediction is expected for a strongly interacting gas.$^{36}$

\begin{figure}
\includegraphics[width=0.8\linewidth]{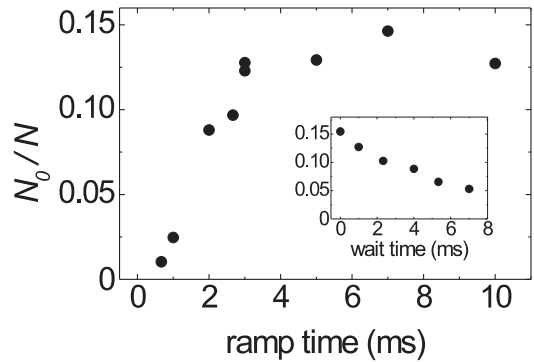}
\caption{Dependence of condensate formation on magnetic-field
sweep rate and measurement of condensate lifetime. We plot the
fraction of condensed molecules versus the time in which the
magnetic field is ramped across the Feshbach resonance from 202.78
G to 201.54 G. The condensate fraction is measured after an
additional waiting time of 1~ms. The initial atom gas temperature
is 0.06 $T_{F}$, the total molecule number is 150,000, and the
final radial trap frequency is 260~Hz. For the full range of ramp
times the number of molecules created remains constant. In the
inset the condensate fraction is plotted versus the wait time
after a 10 ms magnetic-field ramp. The molecule number is not
significantly reduced on this timescale, and the lifetime of the
condensate is instead determined by a heating rate, which we
measure to be 3~$\pm $~1~nK/ms. This heating rate is presumably
due to density dependent inelastic loss processes.}
\end{figure}

We find that the creation of a Bose-Einstein condensate of
molecules requires that the Feshbach resonance be traversed
sufficiently slowly. This is illustrated in Fig. 3, where the
measured condensate fraction is plotted versus the ramp time
across the Feshbach resonance starting with a Fermi gas at a
temperature 0.06 $T_{F}$. Our fastest sweeps result in a much
smaller condensate fraction while the largest condensate fraction
appears for a $B$-field sweep of 3 to 10~ms. For even slower
magnetic field sweeps we find that the condensate fraction slowly
decreases. We attribute this effect to a finite lifetime of the
condensate. Note that the timescale of the experiment is short
compared to the axial trap frequency. Therefore the condensate may
not have global phase coherence in the axial direction.$^{32}$ The
inset of Fig. 3 shows a plot of the lifetime of the condensate.
The observed reduction in condensate fraction is accompanied by
heating of the molecule gas, presumably due to collisional decay
of the molecules into more tightly bound states.
\begin{figure}
\includegraphics[width=0.8\linewidth]{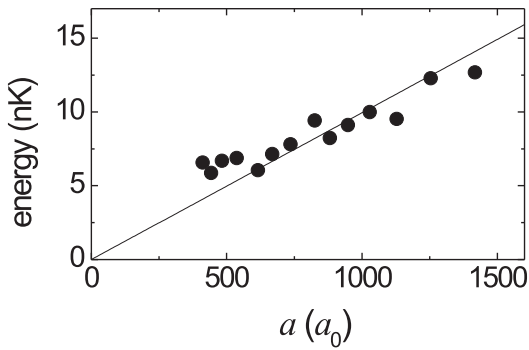}
\caption{Expansion energy per particle for the molecular
condensate versus the interaction strength during expansion. As
the molecular condensate is released from the trap, collisions
convert the mean-field interaction energy into kinetic energy of
the expanding molecules. In this measurement the BEC is created in
a regime for which we calculate the atom-atom scattering length as
$a$ = 3300 $a_{0}$. For time-of-flight expansion the magnetic
field is rapidly changed to different final values. The graph
shows the expansion energy of the condensate fraction determined
from a bimodal fit versus the atom-atom scattering length $a$
corresponding to the magnetic field during expansion. The total
molecule number is 140,000, the magnetic field before expansion is
201.54 G, and we measure the condensate fraction to be 14{\%}. The
line is a linear fit with no offset. We find that the kinetic
energy of the condensate molecules is proportional to the
atom-atom scattering length $a$.}
\end{figure}

Rapidly changing the interaction strength for time-of-flight
expansion of the condensate allows us to measure the interaction
energy in the molecular sample. Figure 4 shows a plot of the
expansion energy of the molecule BEC for various interaction
strengths during time-of-flight expansion. Here the condensate is
created at a fixed interaction strength, and thus the initial peak
density $n_{pk}$ is constant. The data show that the expansion
energy is proportional to the atom-atom scattering length. The
linear dependence suggests that the molecule-molecule scattering
length is proportional to the atom-atom scattering length as
predicted in Ref. 31. In addition we find that the expansion
energy extrapolates to near zero energy for $a$=0. This is
consistent with a Bose-Einstein condensate of molecules. Assuming
the molecule-molecule interaction strength calculated in Ref. 31
this measurement allows us to determine the peak density of the
strongly interacting condensate as $n_{pk}$ = 7 $\times $
10$^{12}$~cm$^{-3}$.
\begin{figure}
\includegraphics[width=0.8\linewidth]{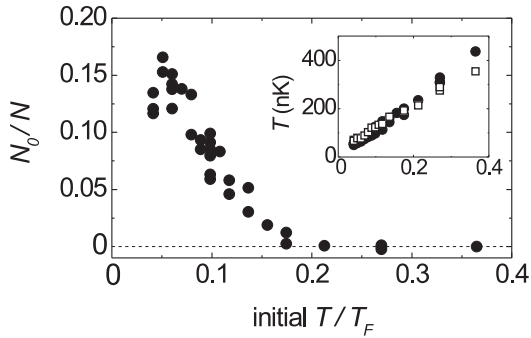}
\caption{Dependence of the condensate fraction and the temperature
of the atom-molecule mixture on the initial scaled temperature
$T/T_{F}$ of the Fermi gas. The condensate fraction is plotted
versus $T/T_{F}$ of the fermionic atoms prior to the magnetic
sweep. In the inset, the temperature of the atoms (open boxes) and
the thermal fraction of the molecules (closed circles) are plotted
versus $T/T_{F}$ before the sweep. This is the same dataset as in
Fig. 2.}
\end{figure}

A fundamental aspect of the experiment is that we start with a
quantum degenerate Fermi gas of atoms. The Bose-Einstein
condensate, which is observed immediately after the magnetic field
is ramped across the Feshbach resonance, therefore requires a
drastic change of the quantum statistical thermodynamics of the
gas. This change is not due to evaporative cooling and the total
number of atoms (adding both free atoms and those bound in
molecular dimers) is conserved by the field sweep. In Figure 5 we
show the dependence of the condensate fraction on the initial
temperature of the Fermi gas. We find that a BEC is formed when
the initial temperature is below 0.17 $T_{F}$. If we assume that
entropy is conserved in the sweep across the Feshbach resonance,
then creating the molecular Bose-Einstein condensate depends on
starting with a Fermi gas at sufficiently low $T/T_{F}$ to give
low initial entropy.$^{15}$ At the onset of BEC our temperature
measurement indicates a modest 40{\%} increase in the total
entropy after the magnetic field sweep, estimated from an ideal
gas model. The inset in Fig. 5 compares the absolute temperature
of atoms and molecules after the magnetic field sweep. For the
molecules the temperature is determined by a fit to the
non-condensate fraction. We find that atoms and molecules are well
thermalized. Note that the atoms and molecules are not in full
chemical equilibrium.$^{12}$ Even though the final binding energy
of the molecules is significantly larger than kinetic energy in
the gas, we only observe conversion efficiencies of up to 88{\%}.
In order to study the reversibility of the ramp across the
Feshbach resonance we have ramped the magnetic field back to the
attractive side of the resonance after creating a molecular
condensate and then measured the temperature of the resulting
Fermi gas. We find that the gas is heated by 27 $\pm $ 7 nK in
this double ramp, independent of the initial temperature.

In conclusion, we have created a Bose-Einstein condensate of
weakly bound molecules starting with a gas of ultracold fermionic
atoms. The molecular BEC has been detected through a bimodal
momentum distribution, and effects of the strong interparticle
interaction have been investigated. The molecular BEC reported
here, which appears on the repulsive side of the Feshbach
resonance, is related in a continuous way to BCS-type fermionic
superfluidity on the attractive side of the resonance. Our
experiment corresponds to the BEC limit, in which superfluidity
occurs due to BEC of essentially local pairs whose binding energy,
$\sim \hbar ^2 / ma^2$, is much larger than the Fermi energy. Here
$m$ is the atomic mass and $2\pi \hbar$ is Planck's constant. The
dimensionless parameter $1/k_F a$, which drives the crossover from
a BCS-superfluid to a molecular BEC$^{3,4}$ is thus positive and
large compared to one. In this regime, the formation of molecules
is clearly separated from their condensation. In contrast, near
the Feshbach resonance where $1/k_F a$ goes through zero, the
system may be described neither by a BEC of molecular dimers nor a
BCS-state of correlated pairs in momentum space, a situation which
has been termed ``resonance superfluidity".$^{6}$ Indeed our
experiment passes through this largely unexplored crossover regime
and with initial temperatures below 0.1 $T_{F}$ the Fermi gas is
well below the predicted critical temperature in the crossover
regime of 0.22 $T_{F}$.$^{9}$ In future work it will be very
exciting to investigate this system on the attractive side of the
resonance and look for evidence of fermionic superfluidity.

We thank L. D. Carr , E. A. Cornell, C. E. Wieman, W. Zwerger, and
I. Bloch for useful discussion and J. Smith for experimental
assistance. This work was supported by NSF and NIST, and C. A. R.
acknowledges support from the Hertz Foundation.

\end{document}